\documentclass[prd,superscriptaddress,twocolumn,nofootinbib,preprintnumbers]{revtex4}
\usepackage{graphicx,amsmath,hyperref}

\begin{document}

\preprint{CP3-16-02, LPSC16007}

\title{Diphoton excess in phenomenological spin-2 resonance scenarios}

\author{Antony Martini}
\affiliation{Centre for Cosmology, Particle Physics and Phenomenology (CP3),
 Universit\'e catholique de Louvain,\\
 B-1348 Louvain-la-Neuve, Belgium}
\author{Kentarou Mawatari}
\affiliation{Laboratoire de Physique Subatomique et de Cosmologie,
Universit\'e Grenoble-Alpes, CNRS/IN2P3, 
53 Avenue des Martyrs, F-38026 Grenoble, France}
\affiliation{Theoretische Natuurkunde and IIHE/ELEM, Vrije Universiteit Brussel,
 and International Solvay Institutes,\\
 Pleinlaan 2, B-1050 Brussels, Belgium}
\author{Dipan Sengupta}
\affiliation{Laboratoire de Physique Subatomique et de Cosmologie,
Universit\'e Grenoble-Alpes, CNRS/IN2P3,
53 Avenue des Martyrs, F-38026 Grenoble, France}


\begin{abstract}
We provide a possible explanation of a 750~GeV diphoton excess recently 
 reported by both the ATLAS and CMS collaborations in the context of
 phenomenological spin-2 resonance scenarios, where the independent
 effective couplings of the resonance with gluons, quarks and photons
 are considered. 
We find a parameter region where the excess can be accounted for without
 conflicting with dijet constraints. 
We also show that the kinematical distributions might help to determine 
 the couplings to gluons and quarks.
\end{abstract}

\maketitle

\noindent
{\it Introduction:}
A heavy spin-2 resonance is one of several well-motivated new physics
candidates, and provides a possible explanation of 750~GeV diphoton excess
recently reported by both the ATLAS~\cite{ATLAS:2015} and 
CMS~\cite{CMS:2015dxe} collaborations with the early LHC Run-II data. 
The CMS analysis~\cite{CMS:2015dxe} considered the Randall--Sundrum (RS)
model~\cite{Randall:1999ee} to constrain the massive graviton as well as
to explain the excess, followed by several theoretical spin-2 
interpretations~\cite{Franceschini:2015kwy,Low:2015qep,Arun:2015ubr,Han:2015cty,Kim:2015ksf,Csaki:2016raa,Buckley:2016mbr}.

In this article, we interpret the excess as a spin-2 resonance that has
independent effective couplings to gluons, quarks and photons as a
minimal phenomenological approach. 
In this model, a massive spin-2 particle is produced via $gg$ and/or
$q\bar q$ initial states and decays into a pair of photons.
We explore the viable parameter region that accounts for the reported 
excess.
The resonance can also decay into a pair of partons, giving dijet final
states at the leading order, which implies that the existing 
8~\cite{Chatrchyan:2013qha,Aad:2014aqa,CMS:2015neg} and  
13~TeV~\cite{Khachatryan:2015dcf,ATLAS:2015nsi} dijet data potentially
exclude a part of the model parameter space. 
This is also taken into account in this study.
Moreover, we investigate kinematical distributions which may give more
information on the couplings to gluons and quarks.\\

\noindent
{\it Model:}
We consider a massive spin-2 particle which couples to the standard
model (SM) gauge and matter fields through their energy--momentum
tensors~\cite{Giudice:1998ck,Han:1998sg} 
\begin{align}
 {\cal L}_{\rm eff} = -\frac{1}{\Lambda}\big[
  \kappa_{\gamma}\,T^{\gamma}_{\mu\nu}
 +\kappa_{g}\,T^{g}_{\mu\nu}
 +\kappa_{q}\,T^{q}_{\mu\nu}
 \big]
 X_2^{\mu\nu}\,,
\label{lag}
\end{align}
where $X_2^{\mu\nu}$ is the spin-2 resonance and
$T^{\gamma,g,q}_{\mu\nu}$ are the energy-momentum tensors; see  
the explicit formula, e.g. in Refs.~\cite{Han:1998sg,Hagiwara:2008jb}. 
Here, for simplicity, we only consider the interactions with photons,
gluons and light quarks. 
While conventional graviton excitations have a universal coupling
strength $\Lambda^{-1}$, where $\Lambda$ is the scale parameter of the
theory, following~\cite{Ellis:2012jv,Englert:2012xt} we introduce the
phenomenological coupling parameters $\kappa_{\gamma}$, $\kappa_g$ and
$\kappa_q$ without assuming any UV models.
We note that the inclusion of interactions with other SM particles such
as leptons, top quarks and weak bosons is straightforward but one should
then consider additional constraints from dilepton, $t\bar t$ and diboson
searches.

We generate the signal events at the parton level by employing the Higgs
Characterisation (HC)~\cite{Artoisenet:2013puc} model,%
\footnote{The model file is publicly available at the
{\sc FeynRules}~\cite{Alloul:2013bka} repository.}
where the Lagrangian~\eqref{lag} was implemented (based
on~\cite{Hagiwara:2008jb,deAquino:2011ix}).%
\footnote{Although the HC model is designed to study the spin--parity
nature of the 125~GeV Higgs boson, one can easily change its mass
$m_{X_2}$ as a parameter.}  
The HC model file is interfaced~\cite{Degrande:2011ua,deAquino:2011ub}
to the {\sc MadGraph5\_aMC@NLO} event generator~\cite{Alwall:2014hca}.\\

\noindent
{\it Analyses (rates):}
In the following we consider three benchmark scenarios:
\begin{itemize}
\item[I:] $\kappa_g\ne0$, $\kappa_q=0$\quad 
 (gluon dominant scenario)\,,\\[-6mm]
\item[II:] $\kappa_g=0$, $\kappa_q\ne0$\quad 
 (quark dominant scenario)\,,\\[-6mm]
\item[III:] $\kappa_g,\,\kappa_q\ne0$\quad
 (mixed scenario)\,,
\end{itemize}
with $\kappa_\gamma\ne0$.
We fix the scale $\Lambda$ at 10~TeV throughout our study.

\begin{figure*}
\center
 \includegraphics[height=5.3cm]{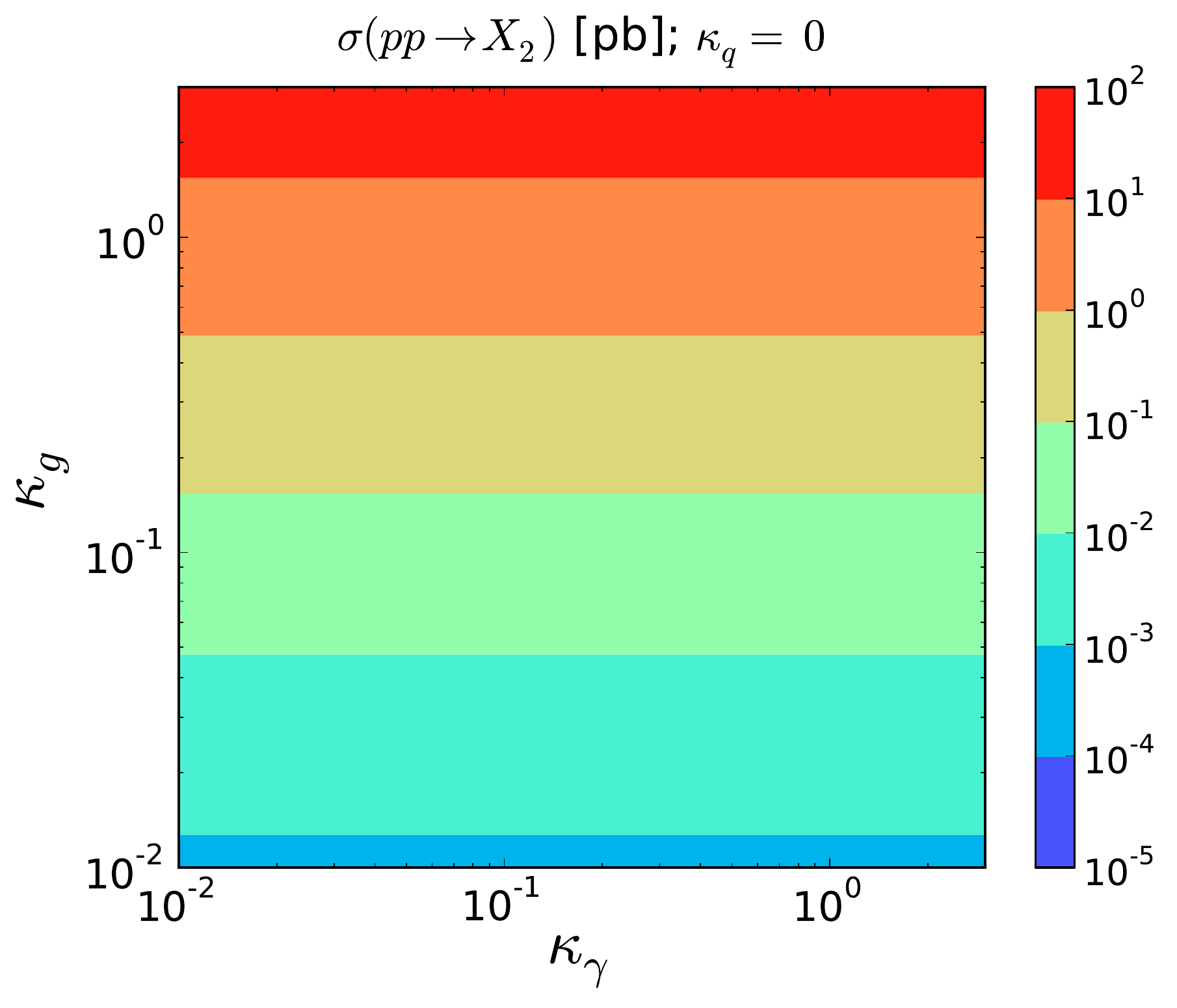}
 \includegraphics[height=5.3cm]{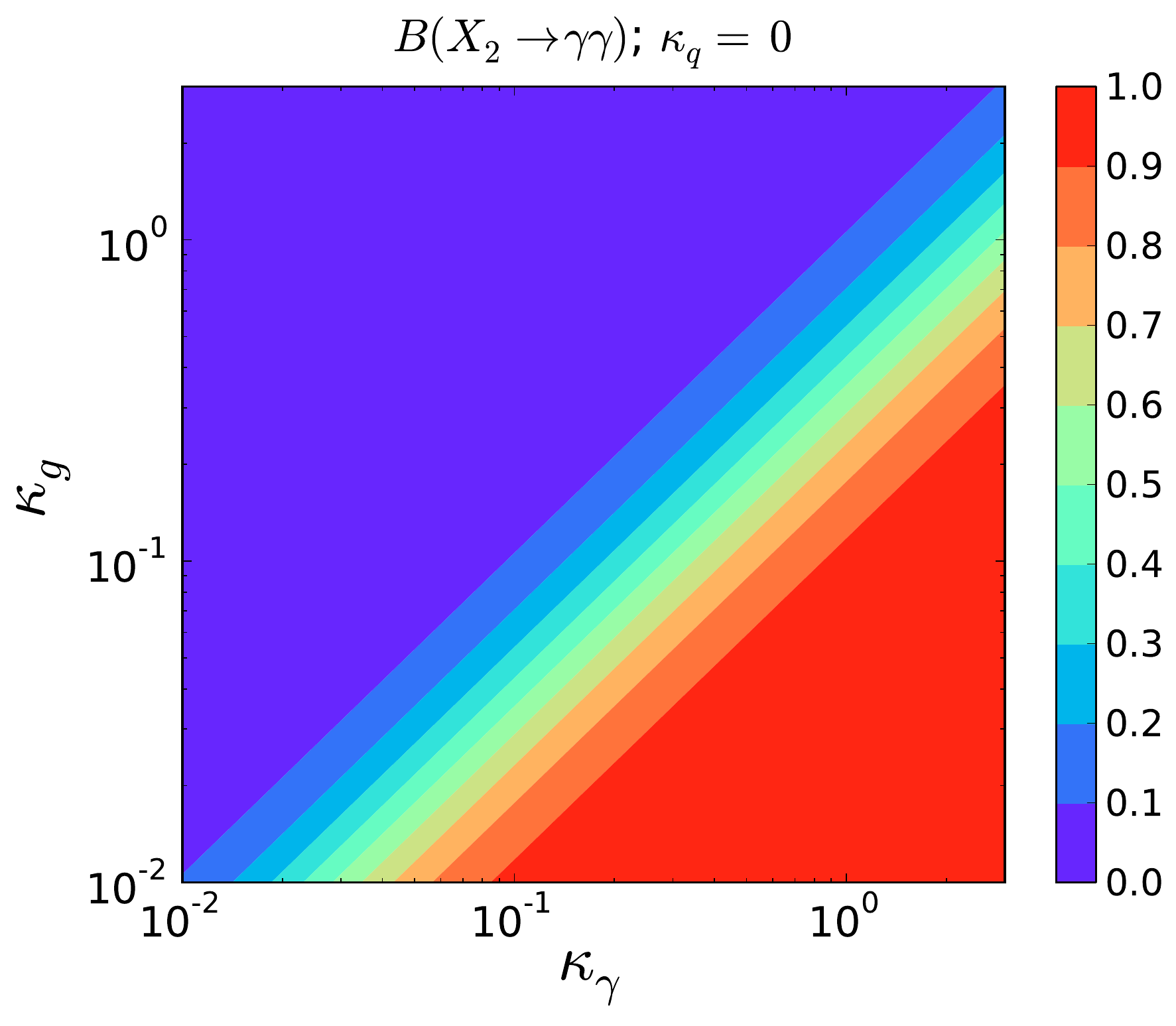}\quad
 \includegraphics[height=5.3cm]{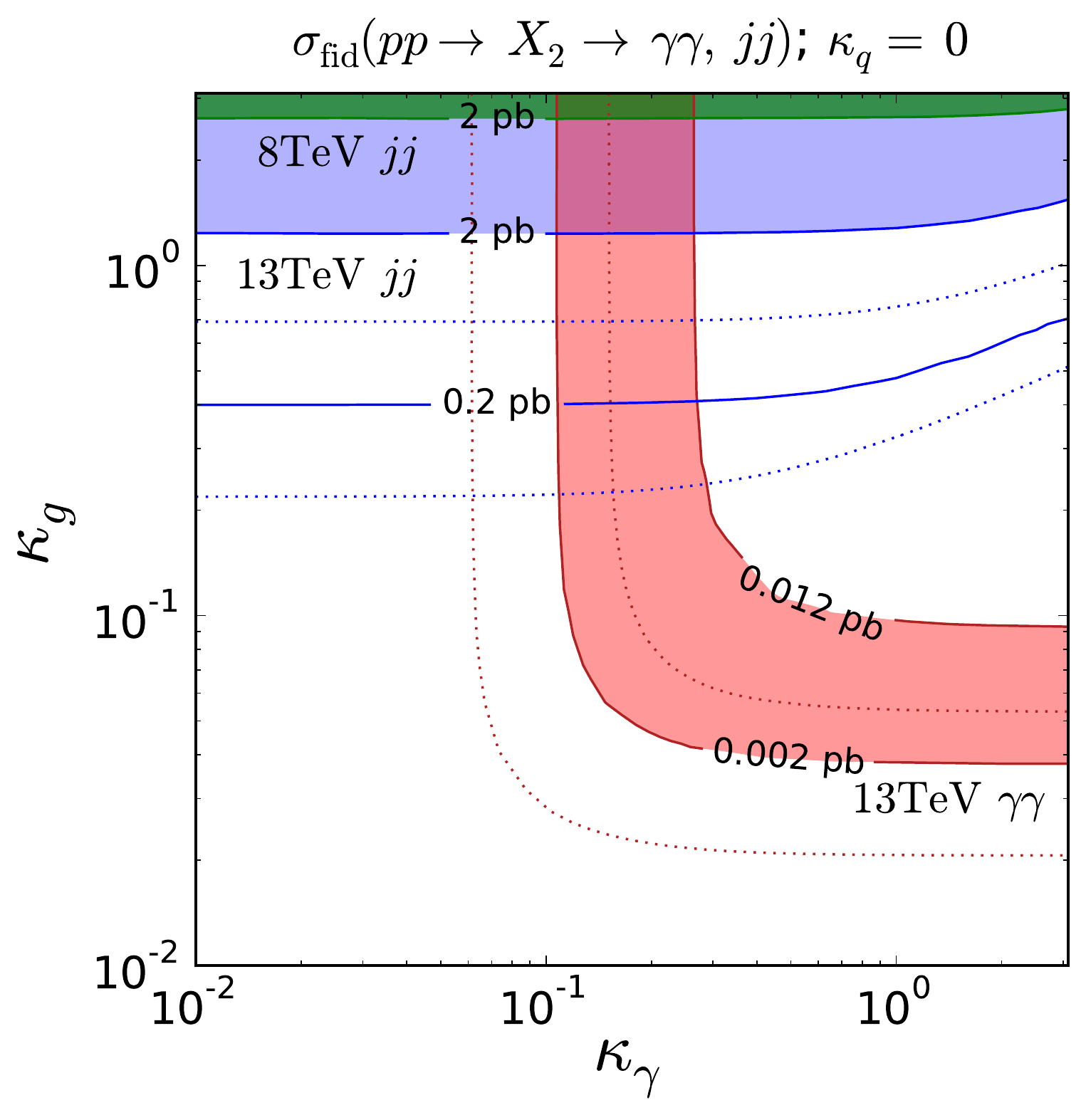}
 \includegraphics[height=5.3cm]{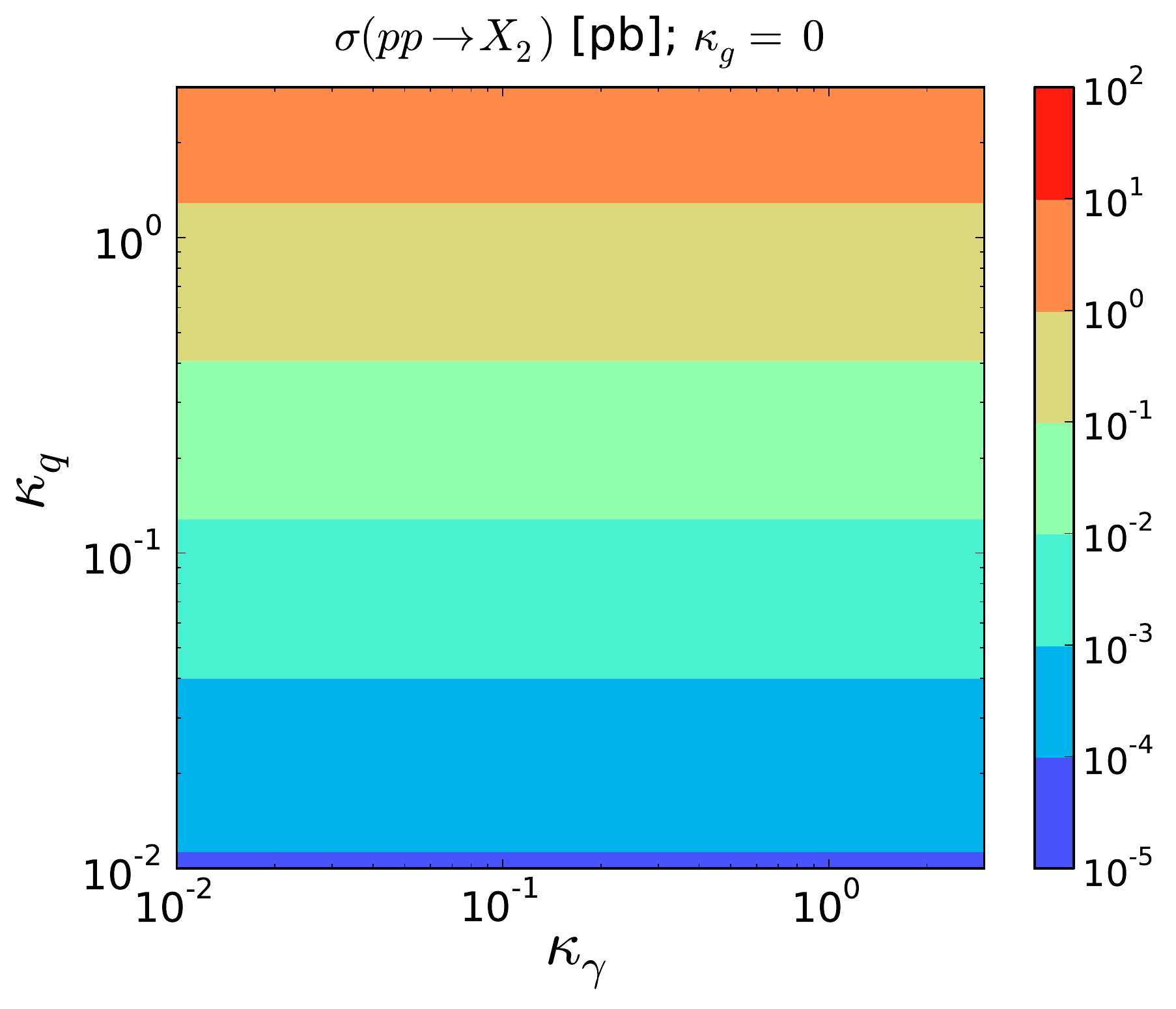}
 \includegraphics[height=5.3cm]{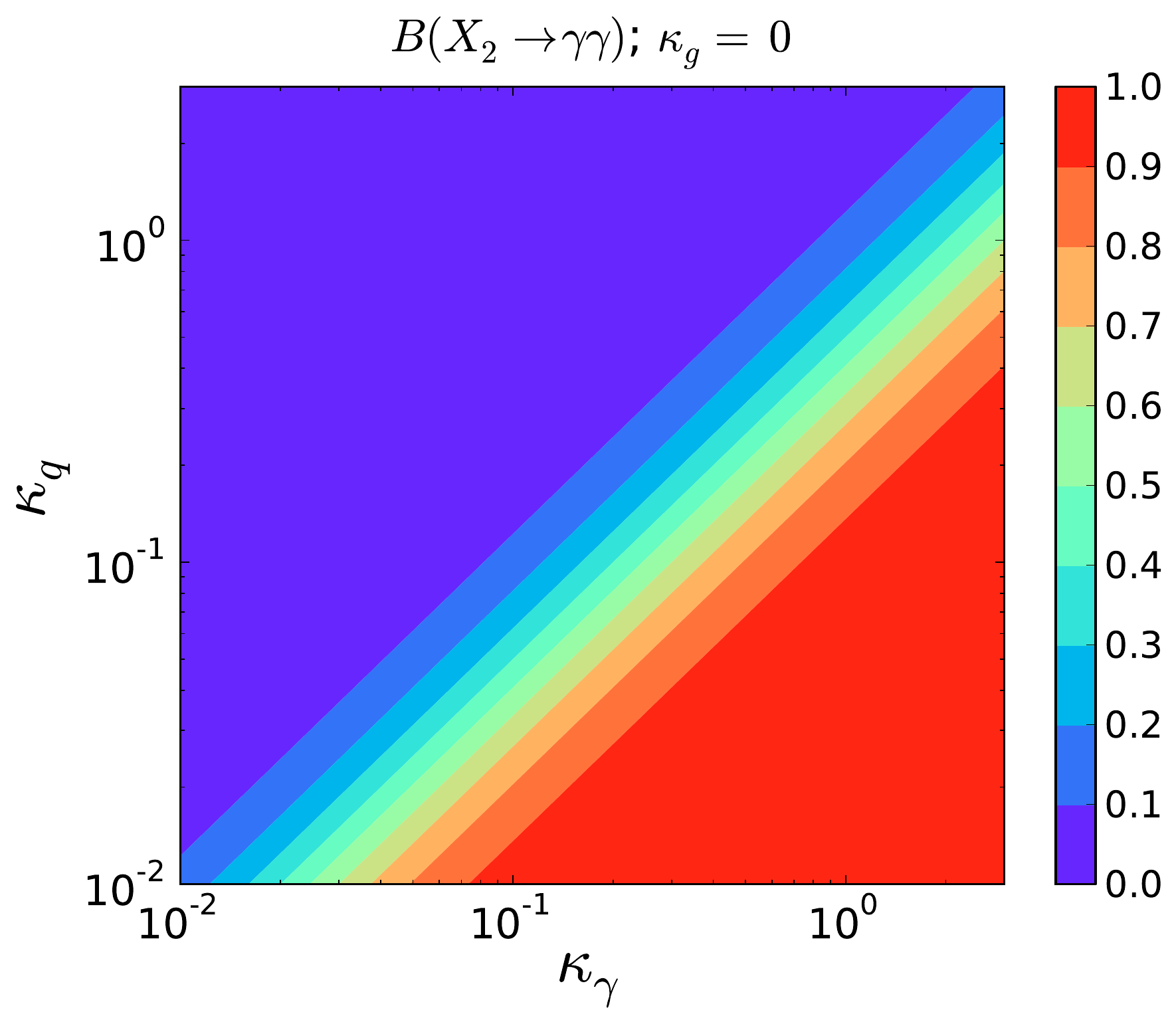}\quad
 \includegraphics[height=5.3cm]{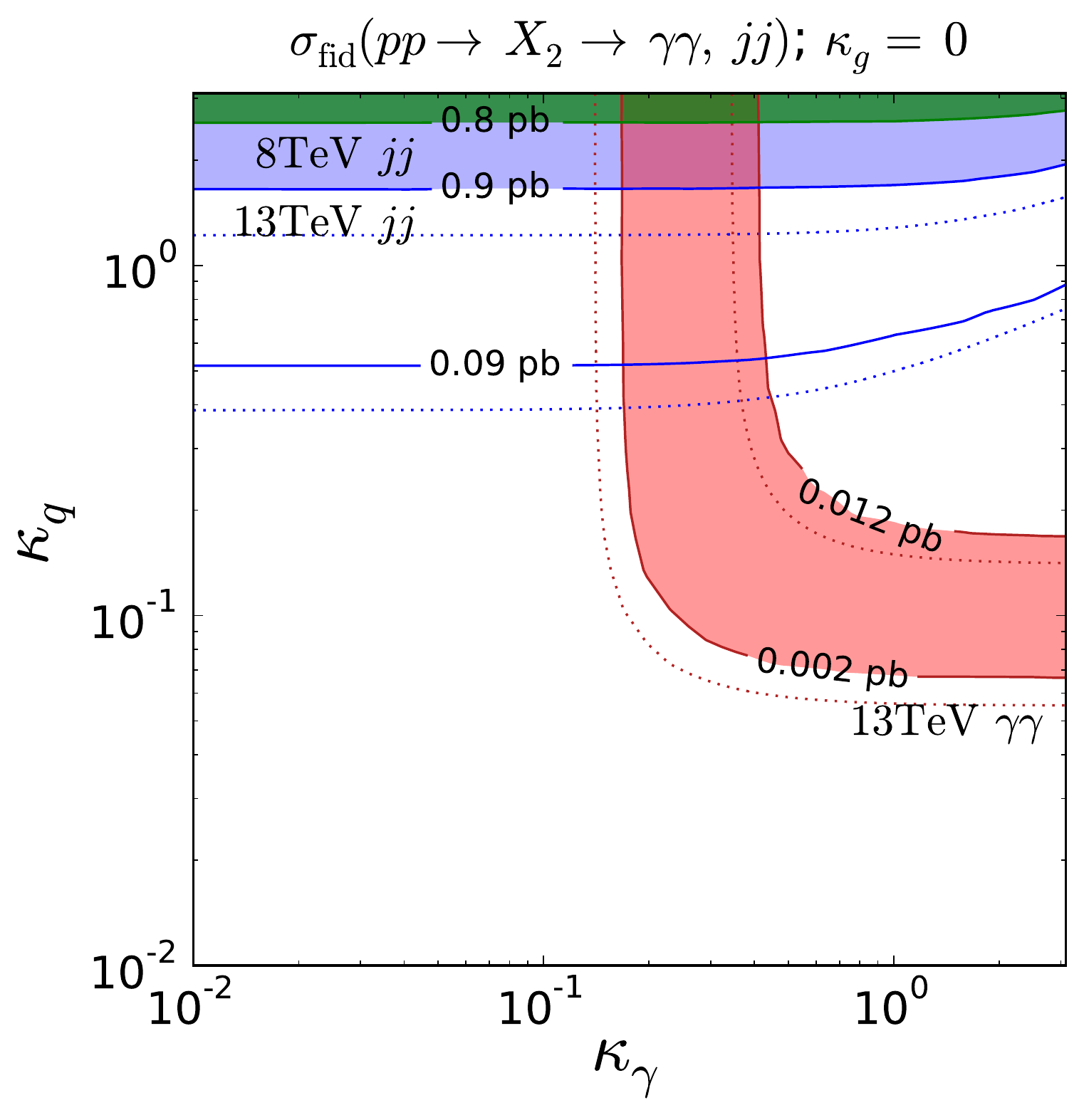}
\caption{
 Left: Total cross section for the spin-2 boson production at the 13~TeV 
 LHC. 
 Center: Diphoton branching ratio.
 Right: Regions explaining the 750~GeV diphoton excess and
 constrained by the dijet data; see the text for details.
 The gluon (quark) dominant scenario I (II) is shown in the top (bottom) 
 panels.   
}
\label{fig:scan}
\end{figure*}

We start from the simple two-dimensional parameter scans, i.e. scenarios
I and II.
Figure~\ref{fig:scan} shows the total cross section for the spin-2 boson
production at the 13~TeV LHC (left), the diphoton branching ratio
(middle), and the region accounting for the 750~GeV diphoton excess
(right) in the $\kappa_\gamma$--$\kappa_g$ (top) and
$\kappa_\gamma$--$\kappa_q$ (bottom) planes. 

The $X_2$ production rate is proportional to $\kappa_g^2$ ($\kappa_q^2$)
in scenario I (II). 
For $\kappa_g=1$ ($\kappa_q=1$) the cross section is 4.2~pb (0.6~pb),
where the NNPDF2.3~\cite{Ball:2012cx} is employed and the factorisation
scale is fixed at $m_{X_2}=750$~GeV. 
For $\kappa_g\sim\kappa_q$ the production rate of the gluon-induced
process is larger than that of the quark-induced one by a factor of
seven due to the larger parton luminosity. 

The diphoton branching ratio depends on both $\kappa_\gamma$ and
$\kappa_{g,q}$, and is determined by the partial widths
\begin{align}
 \Gamma_{\{\gamma\gamma,\,gg,\,qq\}}
  =\frac{m_{X_2}^3}{80\pi\Lambda^2}
   \big\{\kappa_\gamma^2,\,8\kappa_g^2,\,6\kappa_q^2 \big\}\,,
\label{width}
\end{align}
where four quark flavours are considered.
{\sc MadWidth}~\cite{Alwall:2014bza} provides the above partial decay 
rates for each parameter point.
The $X_2$ total width is 1.4 (1.1)~GeV for
$\kappa_\gamma=\kappa_{g(q)}=3$; i.e. the $X_2$ has a narrow width
for the entire parameter space in our scenarios. 
We will briefly discuss the broad resonance scenario in the summary
section. 

In the right plots in Fig.~\ref{fig:scan}, the red shaded region can fit 
the ATLAS and CMS 
diphoton excess (2--12~fb)~\cite{ATLAS:2015,CMS:2015dxe}, where, in
addition to the minimal cuts $p_T^\gamma>25$~GeV and
$|\eta^\gamma|<2.37$, the fiducial cut~\cite{ATLAS:2015} 
\begin{align}
 p_T^\gamma>300~{\rm GeV}
\end{align}
is imposed. 
The red dotted lines refer to the same region but without any cuts,
i.e. derived by the simple 
$\sigma(pp\to X_2)\times B(X_2\to\gamma\gamma)$ calculation.
One can easily understand the shape of the region from 
(the left plot) $\times$ (the middle plot).
Analytically, $\sigma\times B_{\gamma\gamma}\propto\kappa_\gamma^2$ for 
$\kappa_\gamma\ll\kappa_{g,q}$, while
$\sigma\times B_{\gamma\gamma}\propto\kappa_{g,q}^2$ for
$\kappa_\gamma\gg\kappa_{g,q}$. 
For the quantitative estimation, on the other hand, we find that the
effect of the fiducial cut is significant. 
We checked that the effect of the minimal cuts is small.
It should be noted that the effect of the fiducial cut is larger for the
gluon case (I) than that for the quark one (II).
This can be explained by the difference of the kinematical distributions
between the two cases, which will be discussed later.

As mentioned in the Introduction, the existing dijet data may constrain
the allowed parameter space. 
The green and blue shaded regions in the right plots of
Fig.~\ref{fig:scan} present the exclusion by the dijet analyses at 
$\sqrt{s}=8$~\cite{Chatrchyan:2013qha,Aad:2014aqa,CMS:2015neg} and 
13~TeV~\cite{Khachatryan:2015dcf,ATLAS:2015nsi}, respectively. 
To evaluate the dijet constraints, in addition to the minimal cuts
$p_T^j>30$~GeV and $|\eta^j|<2.5$, the selection
cut~\cite{CMS:2015neg,Khachatryan:2015dcf}   
\begin{align}
 |\Delta\eta^{jj}|<1.3
\end{align}
is imposed. 
We take the 95\% confidence level (CL) upper limits on the cross section
at $\sqrt{s}=8$~TeV from the CMS analysis~\cite{CMS:2015neg}: 2.0
(0.8)~pb for the  
$gg$ ($q\bar q$) type dijet resonance.
For the 13~TeV dijet constraint, on the other hand, there is no analysis
below $m_{jj}<1$~TeV yet~\cite{Khachatryan:2015dcf,ATLAS:2015nsi}, and
hence we simply extrapolate the CMS limits at the 1.5~TeV resonance
mass, i.e. 2.0 (0.9)~pb for the $gg$ ($q\bar q$) type
resonance~\cite{Khachatryan:2015dcf}, as a conservative constraint.   
We also show the ten-time stronger limits (0.2~pb for $gg$ and 0.09~pb
for $q\bar q$) at $\sqrt{s}=13$~TeV as a future constraint.
Similar to the diphoton case, the blue dotted lines show the 13~TeV
dijet constraints derived by $\sigma\times B_{jj}$ without any cuts.
For the dijet case, $\sigma\times B_{jj}\propto\kappa_{g,q}^2$ for 
$\kappa_\gamma\ll\kappa_{g,q}$, while
$\sigma\times B_{jj}\propto\kappa_{g,q}^4/\kappa_\gamma^2$ for
$\kappa_\gamma\gg\kappa_{g,q}$.

We conclude that the current (and near future) dijet constraints are not
strong enough to exclude the parameter region accounting for the
diphoton excess for both the $gg$ and $q\bar q$ scenarios. 
We note that the $\kappa_{g(q)}=\kappa_\gamma=0.25\ (0.4)$ case gives 
$\sigma_{\rm fid}(\gamma\gamma)\sim10$~fb.
Since the $\kappa_g$--$\kappa_q$ mixed scenario III is estimated by the
linear combination of the two scenarios I and II, we can always adjust
the parameters to fit the diphoton excess without conflicting the dijet 
constraints.  
However, the information of the total rate is not enough to determine a
unique ($\kappa_g, \kappa_q$) solution.
We note that since the effect of the fiducial cuts is sizeable,
a more precise analysis should include, e.g., next-to-leading-order
(NLO) corrections in QCD~\cite{Kumar:2009nn}, effects of additional
partons in the final state~\cite{Artoisenet:2013puc}, hadronisation, and
detector response. 
Such a study will be reported elsewhere.\\

\noindent
{\it Analyses (distributions):}
We now turn to the possibility of determining the values of $\kappa_g$
and $\kappa_q$ for the mixed scenario III in the diphoton events.
Since angular dependence at the partonic centre-of-mass frame for a
spin-2 particle production is different between $gg$ and $q\bar q$
initial states~\cite{Allanach:2002gn}, kinematical distributions may be
able to provide additional information. 

\begin{figure}
\center
 \includegraphics[width=1\columnwidth]{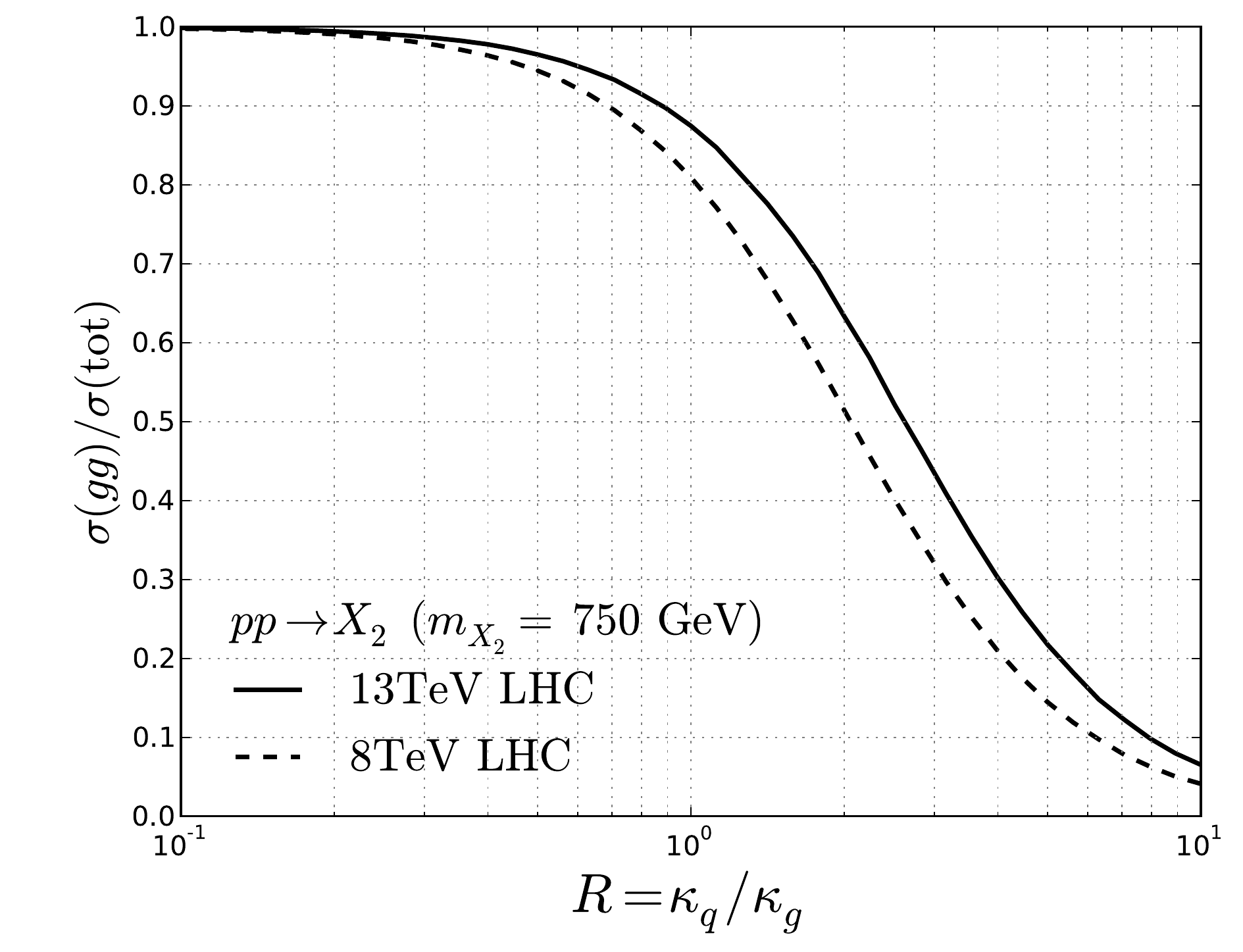}
\caption{Contributions of $gg$ and $q\bar q$ subprocesses to 750~GeV
 spin-2 particle production at the LHC as a function of 
 $R=\kappa_q/\kappa_g$.}
\label{fig:rplot}
\end{figure}

Figure~\ref{fig:rplot} shows the individual contributions of $gg$ and
$q\bar q$ initial states in $pp\to X_2$ production at the 13~TeV LHC as
a function of the ratio between $\kappa_q$ and $\kappa_g$, 
$R\equiv\kappa_q/\kappa_g$.
As a reference, the 8~TeV LHC case is also shown by a dashed line.
As mentioned above, the gluon-induced process is dominant at $R=1$,
i.e. the universal coupling case. 
To present the kinematical distributions below, we choose four benchmark
points as $R=0.1, 1.0, 2.5, 10$, which give 99\%, 87\%, 52\%, 7\% $gg$
contributions, respectively.  
For a certain $R$, we can always find the $\kappa$ parameters to satisfy
both the excess and the constraints.

\begin{figure}
\center
 \includegraphics[width=1\columnwidth]{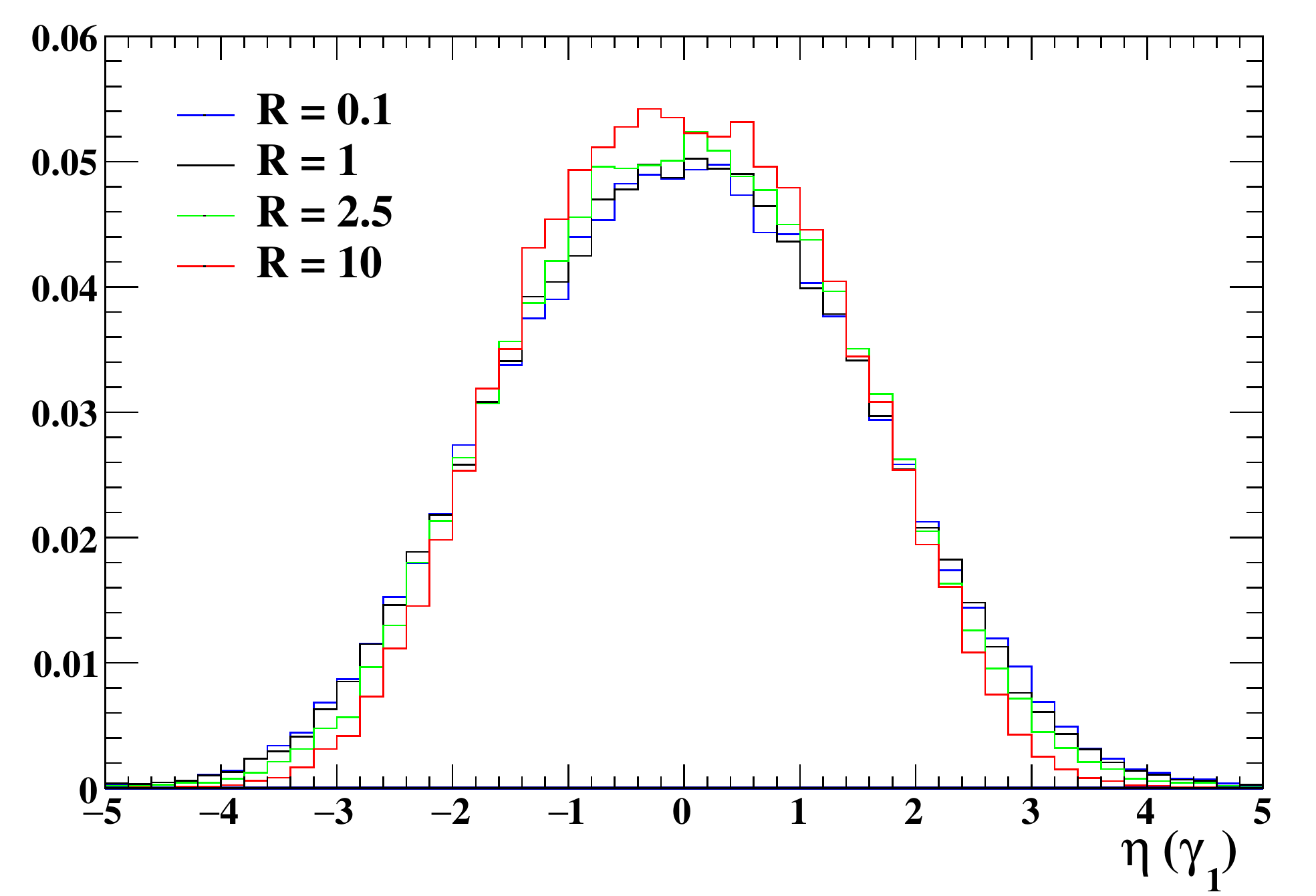}
 \includegraphics[width=1\columnwidth]{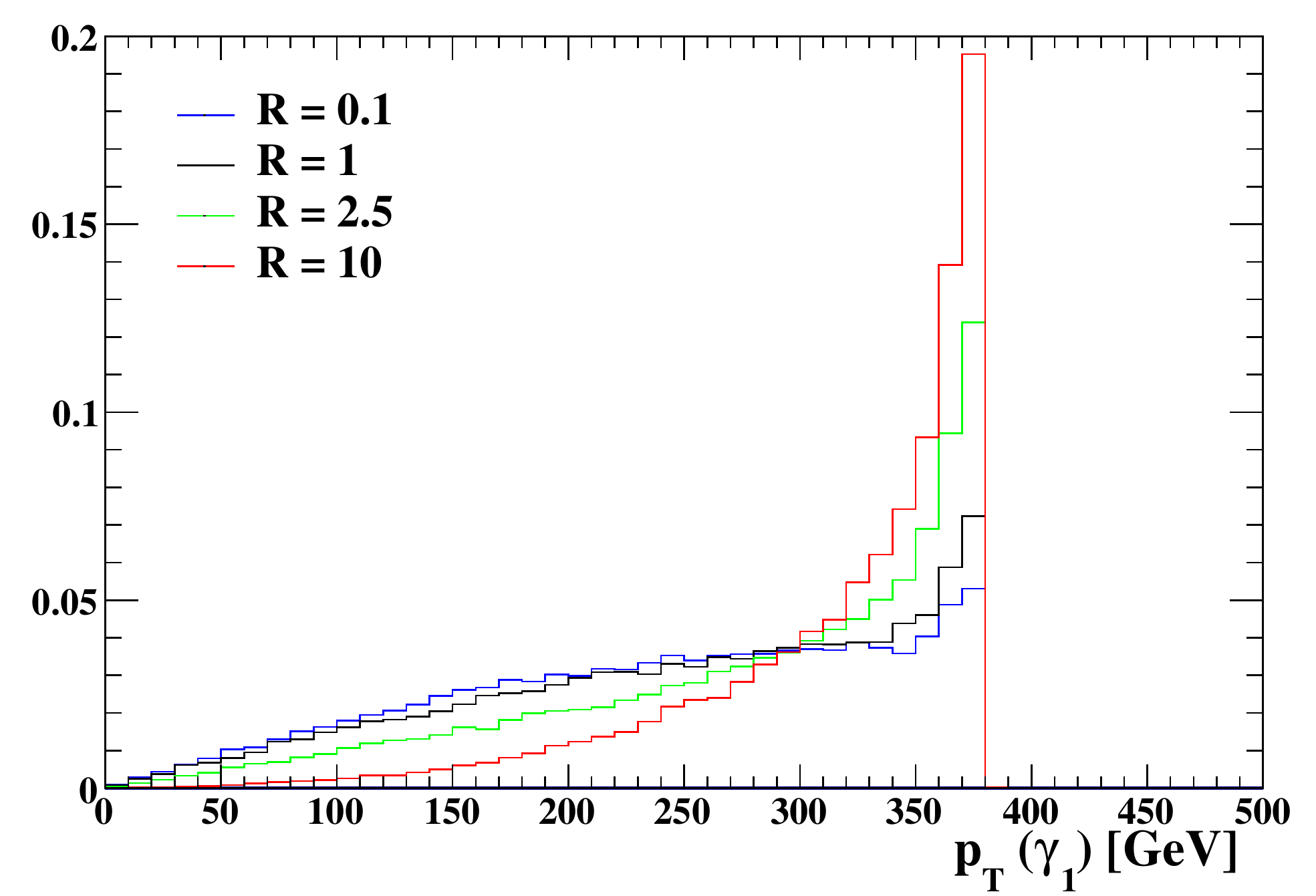}
\caption{
 Normalised photon pseudorapidity (top) and transverse momentum (bottom)
 distributions in $pp\to X_2\to\gamma\gamma$ process at
 $\sqrt{s}=13$~TeV for different $R$ values, without any kinematical
 cuts.}  
\label{fig:dis}
\end{figure}

Figure~\ref{fig:dis} shows the normalised $\eta$ and $p_T$ distributions 
of the photon for the above benchmark points.
Comparing with the gluon-induced process ($R=0.1$), $q\bar q$
annihilation ($R=10$) produces the photons in the more central and
higher-$p_T$ regions. 
These kinematical features can explain the large difference of the
signal acceptance between scenarios I and II, observed in
Fig.~\ref{fig:scan}(right).  
Although the fiducial cut diminishes the distinction between the different
mixed cases, we could still observe the difference once we have enough
data. 
A variation of the fiducial volumes can help to resolve the $\kappa_g$
and $\kappa_q$ values. 
One caveat is that nonuniversal couplings to gluons and quarks, 
$\kappa_g\ne\kappa_q$, give rise to a unitarity-violating behaviour at
the NLO in QCD, especially in the $p_T^{X_2}>m_{X_2}$ 
region~\cite{Artoisenet:2013puc}.
We note, however, that we can always find the region explaining the
excess under the $\kappa_g=\kappa_q$ condition; e.g. 
$\kappa_g=\kappa_q=0.1$ with $\kappa_\gamma=0.4$, providing
$\sigma_{\rm fid}(\gamma\gamma)\sim10$~fb.\\

\noindent
{\it Summary and discussion:}
We interpreted the 750~GeV diphoton anomaly recently reported at the LHC
as a spin-2 narrow resonance which only couples to photons, gluons and
light quarks. 
We introduced the three independent coupling parameters,
$\kappa_\gamma$, $\kappa_g$ and $\kappa_q$, and found the viable
parameter regions explaining the diphoton excess without conflicting the
dijet constraints. 
The diphoton kinematical distributions are distinctive between the $gg$
and $q\bar q$ subprocesses, which leads to different signal acceptances
and provides a possibility to determine the parameters uniquely. 

Although we only considered the spin-2 couplings to the SM particles
relevant to the diphoton and dijet analyses as a minimal
phenomenological framework without assuming any UV models,  
extensions of the model are straightforward, but should include more
constraints such as dilepton, diboson, and $t\bar t$ final states.

Before closing, we mention the broader resonance case, which is
indicated by the ATLAS data~\cite{ATLAS:2015}.
Our viable parameter region only provides a narrow width even after
including all the SM decay channels.
In passing, $\Gamma_{X_2}\sim45$~GeV requires $\Lambda\sim 950$~GeV in
the RS model with a universal coupling, which is excluded by the dijet
data. 
Therefore, other decay channels such as dark matter~\cite{Han:2015cty}
are necessary to have such a wider resonance.\\

\noindent
{\it Acknowledgements:}
We would like to thank Sabine Kraml and Fabio Maltoni for valuable
comments on the manuscript.

The work of A.\,M. is supported by the IISN ``MadGraph'' convention
4.4511.10 and the IISN ``Fundamental interactions'' convention
4.4517.08.
The work of K.\,M. is supported by the Theory-LHC-France initiative of the
CNRS (INP/IN2P3).
The work of D.\,S. is supported by the French ANR project DMAstroLHC,
ANR-12-BS05-0006.

\bibliography{bibdiphoton}

\begin{thebibliography}{30}
\expandafter\ifx\csname natexlab\endcsname\relax\def\natexlab#1{#1}\fi
\expandafter\ifx\csname bibnamefont\endcsname\relax
  \def\bibnamefont#1{#1}\fi
\expandafter\ifx\csname bibfnamefont\endcsname\relax
  \def\bibfnamefont#1{#1}\fi
\expandafter\ifx\csname citenamefont\endcsname\relax
  \def\citenamefont#1{#1}\fi
\expandafter\ifx\csname url\endcsname\relax
  \def\url#1{\texttt{#1}}\fi
\expandafter\ifx\csname urlprefix\endcsname\relax\def\urlprefix{URL }\fi
\providecommand{\bibinfo}[2]{#2}
\providecommand{\eprint}[2][]{\url{#2}}

\bibitem[{\citenamefont{ATLAS}(2015)}]{ATLAS:2015}
\bibinfo{author}{\bibnamefont{ATLAS}} (\bibinfo{year}{2015}),
  \eprint{ATLAS-CONF-2015-081}.

\bibitem[{\citenamefont{CMS}(2015{\natexlab{a}})}]{CMS:2015dxe}
\bibinfo{author}{\bibnamefont{CMS}} (\bibinfo{year}{2015}{\natexlab{a}}),
  \eprint{CMS-PAS-EXO-15-004}.

\bibitem[{\citenamefont{Randall and Sundrum}(1999)}]{Randall:1999ee}
\bibinfo{author}{\bibfnamefont{L.}~\bibnamefont{Randall}} \bibnamefont{and}
  \bibinfo{author}{\bibfnamefont{R.}~\bibnamefont{Sundrum}},
  \bibinfo{journal}{Phys. Rev. Lett.} \textbf{\bibinfo{volume}{83}},
  \bibinfo{pages}{3370} (\bibinfo{year}{1999}), \eprint{hep-ph/9905221}.

\bibitem[{\citenamefont{Franceschini et~al.}(2015)\citenamefont{Franceschini,
  Giudice, Kamenik, McCullough, Pomarol, Rattazzi, Redi, Riva, Strumia, and
  Torre}}]{Franceschini:2015kwy}
\bibinfo{author}{\bibfnamefont{R.}~\bibnamefont{Franceschini}},
  \bibinfo{author}{\bibfnamefont{G.~F.} \bibnamefont{Giudice}},
  \bibinfo{author}{\bibfnamefont{J.~F.} \bibnamefont{Kamenik}},
  \bibinfo{author}{\bibfnamefont{M.}~\bibnamefont{McCullough}},
  \bibinfo{author}{\bibfnamefont{A.}~\bibnamefont{Pomarol}},
  \bibinfo{author}{\bibfnamefont{R.}~\bibnamefont{Rattazzi}},
  \bibinfo{author}{\bibfnamefont{M.}~\bibnamefont{Redi}},
  \bibinfo{author}{\bibfnamefont{F.}~\bibnamefont{Riva}},
  \bibinfo{author}{\bibfnamefont{A.}~\bibnamefont{Strumia}}, \bibnamefont{and}
  \bibinfo{author}{\bibfnamefont{R.}~\bibnamefont{Torre}}
  (\bibinfo{year}{2015}), \eprint{1512.04933}.

\bibitem[{\citenamefont{Low et~al.}(2016)\citenamefont{Low, Tesi, and
  Wang}}]{Low:2015qep}
\bibinfo{author}{\bibfnamefont{M.}~\bibnamefont{Low}},
  \bibinfo{author}{\bibfnamefont{A.}~\bibnamefont{Tesi}}, \bibnamefont{and}
  \bibinfo{author}{\bibfnamefont{L.-T.} \bibnamefont{Wang}},
  \bibinfo{journal}{JHEP} \textbf{\bibinfo{volume}{03}}, \bibinfo{pages}{108}
  (\bibinfo{year}{2016}), \eprint{1512.05328}.

\bibitem[{\citenamefont{Arun and Saha}(2015)}]{Arun:2015ubr}
\bibinfo{author}{\bibfnamefont{M.~T.} \bibnamefont{Arun}} \bibnamefont{and}
  \bibinfo{author}{\bibfnamefont{P.}~\bibnamefont{Saha}}
  (\bibinfo{year}{2015}), \eprint{1512.06335}.

\bibitem[{\citenamefont{Han et~al.}(2016)\citenamefont{Han, Lee, Park, and
  Sanz}}]{Han:2015cty}
\bibinfo{author}{\bibfnamefont{C.}~\bibnamefont{Han}},
  \bibinfo{author}{\bibfnamefont{H.~M.} \bibnamefont{Lee}},
  \bibinfo{author}{\bibfnamefont{M.}~\bibnamefont{Park}}, \bibnamefont{and}
  \bibinfo{author}{\bibfnamefont{V.}~\bibnamefont{Sanz}},
  \bibinfo{journal}{Phys. Lett.} \textbf{\bibinfo{volume}{B755}},
  \bibinfo{pages}{371} (\bibinfo{year}{2016}), \eprint{1512.06376}.

\bibitem[{\citenamefont{Kim et~al.}(2015)\citenamefont{Kim, Rolbiecki, and
  de~Austri}}]{Kim:2015ksf}
\bibinfo{author}{\bibfnamefont{J.~S.} \bibnamefont{Kim}},
  \bibinfo{author}{\bibfnamefont{K.}~\bibnamefont{Rolbiecki}},
  \bibnamefont{and} \bibinfo{author}{\bibfnamefont{R.~R.}
  \bibnamefont{de~Austri}} (\bibinfo{year}{2015}), \eprint{1512.06797}.

\bibitem[{\citenamefont{Csaki et~al.}(2016)\citenamefont{Csaki, Hubisz,
  Lombardo, and Terning}}]{Csaki:2016raa}
\bibinfo{author}{\bibfnamefont{C.}~\bibnamefont{Csaki}},
  \bibinfo{author}{\bibfnamefont{J.}~\bibnamefont{Hubisz}},
  \bibinfo{author}{\bibfnamefont{S.}~\bibnamefont{Lombardo}}, \bibnamefont{and}
  \bibinfo{author}{\bibfnamefont{J.}~\bibnamefont{Terning}}
  (\bibinfo{year}{2016}), \eprint{1601.00638}.

\bibitem[{\citenamefont{Buckley}(2016)}]{Buckley:2016mbr}
\bibinfo{author}{\bibfnamefont{M.~R.} \bibnamefont{Buckley}}
  (\bibinfo{year}{2016}), \eprint{1601.04751}.

\bibitem[{\citenamefont{Chatrchyan et~al.}(2013)}]{Chatrchyan:2013qha}
\bibinfo{author}{\bibfnamefont{S.}~\bibnamefont{Chatrchyan}}
  \bibnamefont{et~al.} (\bibinfo{collaboration}{CMS}), \bibinfo{journal}{Phys.
  Rev.} \textbf{\bibinfo{volume}{D87}}, \bibinfo{pages}{114015}
  (\bibinfo{year}{2013}), \eprint{1302.4794}.

\bibitem[{\citenamefont{Aad et~al.}(2015)}]{Aad:2014aqa}
\bibinfo{author}{\bibfnamefont{G.}~\bibnamefont{Aad}} \bibnamefont{et~al.}
  (\bibinfo{collaboration}{ATLAS}), \bibinfo{journal}{Phys. Rev.}
  \textbf{\bibinfo{volume}{D91}}, \bibinfo{pages}{052007}
  (\bibinfo{year}{2015}), \eprint{1407.1376}.

\bibitem[{\citenamefont{CMS}(2015{\natexlab{b}})}]{CMS:2015neg}
\bibinfo{author}{\bibnamefont{CMS}} (\bibinfo{year}{2015}{\natexlab{b}}),
  \eprint{CMS-PAS-EXO-14-005}.

\bibitem[{\citenamefont{Khachatryan et~al.}(2016)}]{Khachatryan:2015dcf}
\bibinfo{author}{\bibfnamefont{V.}~\bibnamefont{Khachatryan}}
  \bibnamefont{et~al.} (\bibinfo{collaboration}{CMS}), \bibinfo{journal}{Phys.
  Rev. Lett.} \textbf{\bibinfo{volume}{116}}, \bibinfo{pages}{071801}
  (\bibinfo{year}{2016}), \eprint{1512.01224}.

\bibitem[{\citenamefont{Aad et~al.}(2016)}]{ATLAS:2015nsi}
\bibinfo{author}{\bibfnamefont{G.}~\bibnamefont{Aad}} \bibnamefont{et~al.}
  (\bibinfo{collaboration}{ATLAS}), \bibinfo{journal}{Phys. Lett.}
  \textbf{\bibinfo{volume}{B754}}, \bibinfo{pages}{302} (\bibinfo{year}{2016}),
  \eprint{1512.01530}.

\bibitem[{\citenamefont{Giudice et~al.}(1999)\citenamefont{Giudice, Rattazzi,
  and Wells}}]{Giudice:1998ck}
\bibinfo{author}{\bibfnamefont{G.~F.} \bibnamefont{Giudice}},
  \bibinfo{author}{\bibfnamefont{R.}~\bibnamefont{Rattazzi}}, \bibnamefont{and}
  \bibinfo{author}{\bibfnamefont{J.~D.} \bibnamefont{Wells}},
  \bibinfo{journal}{Nucl. Phys.} \textbf{\bibinfo{volume}{B544}},
  \bibinfo{pages}{3} (\bibinfo{year}{1999}), \eprint{hep-ph/9811291}.

\bibitem[{\citenamefont{Han et~al.}(1999)\citenamefont{Han, Lykken, and
  Zhang}}]{Han:1998sg}
\bibinfo{author}{\bibfnamefont{T.}~\bibnamefont{Han}},
  \bibinfo{author}{\bibfnamefont{J.~D.} \bibnamefont{Lykken}},
  \bibnamefont{and} \bibinfo{author}{\bibfnamefont{R.-J.} \bibnamefont{Zhang}},
  \bibinfo{journal}{Phys. Rev.} \textbf{\bibinfo{volume}{D59}},
  \bibinfo{pages}{105006} (\bibinfo{year}{1999}), \eprint{hep-ph/9811350}.

\bibitem[{\citenamefont{Hagiwara et~al.}(2008)\citenamefont{Hagiwara, Kanzaki,
  Li, and Mawatari}}]{Hagiwara:2008jb}
\bibinfo{author}{\bibfnamefont{K.}~\bibnamefont{Hagiwara}},
  \bibinfo{author}{\bibfnamefont{J.}~\bibnamefont{Kanzaki}},
  \bibinfo{author}{\bibfnamefont{Q.}~\bibnamefont{Li}}, \bibnamefont{and}
  \bibinfo{author}{\bibfnamefont{K.}~\bibnamefont{Mawatari}},
  \bibinfo{journal}{Eur. Phys. J.} \textbf{\bibinfo{volume}{C56}},
  \bibinfo{pages}{435} (\bibinfo{year}{2008}), \eprint{0805.2554}.

\bibitem[{\citenamefont{Ellis et~al.}(2013)\citenamefont{Ellis, Fok, Hwang,
  Sanz, and You}}]{Ellis:2012jv}
\bibinfo{author}{\bibfnamefont{J.}~\bibnamefont{Ellis}},
  \bibinfo{author}{\bibfnamefont{R.}~\bibnamefont{Fok}},
  \bibinfo{author}{\bibfnamefont{D.~S.} \bibnamefont{Hwang}},
  \bibinfo{author}{\bibfnamefont{V.}~\bibnamefont{Sanz}}, \bibnamefont{and}
  \bibinfo{author}{\bibfnamefont{T.}~\bibnamefont{You}}, \bibinfo{journal}{Eur.
  Phys. J.} \textbf{\bibinfo{volume}{C73}}, \bibinfo{pages}{2488}
  (\bibinfo{year}{2013}), \eprint{1210.5229}.

\bibitem[{\citenamefont{Englert et~al.}(2013)\citenamefont{Englert,
  Goncalves-Netto, Mawatari, and Plehn}}]{Englert:2012xt}
\bibinfo{author}{\bibfnamefont{C.}~\bibnamefont{Englert}},
  \bibinfo{author}{\bibfnamefont{D.}~\bibnamefont{Goncalves-Netto}},
  \bibinfo{author}{\bibfnamefont{K.}~\bibnamefont{Mawatari}}, \bibnamefont{and}
  \bibinfo{author}{\bibfnamefont{T.}~\bibnamefont{Plehn}},
  \bibinfo{journal}{JHEP} \textbf{\bibinfo{volume}{01}}, \bibinfo{pages}{148}
  (\bibinfo{year}{2013}), \eprint{1212.0843}.

\bibitem[{\citenamefont{Artoisenet et~al.}(2013)}]{Artoisenet:2013puc}
\bibinfo{author}{\bibfnamefont{P.}~\bibnamefont{Artoisenet}}
  \bibnamefont{et~al.}, \bibinfo{journal}{JHEP} \textbf{\bibinfo{volume}{11}},
  \bibinfo{pages}{043} (\bibinfo{year}{2013}), \eprint{1306.6464}.

\bibitem[{\citenamefont{Alloul et~al.}(2014)\citenamefont{Alloul, Christensen,
  Degrande, Duhr, and Fuks}}]{Alloul:2013bka}
\bibinfo{author}{\bibfnamefont{A.}~\bibnamefont{Alloul}},
  \bibinfo{author}{\bibfnamefont{N.~D.} \bibnamefont{Christensen}},
  \bibinfo{author}{\bibfnamefont{C.}~\bibnamefont{Degrande}},
  \bibinfo{author}{\bibfnamefont{C.}~\bibnamefont{Duhr}}, \bibnamefont{and}
  \bibinfo{author}{\bibfnamefont{B.}~\bibnamefont{Fuks}},
  \bibinfo{journal}{Comput.Phys.Commun.} \textbf{\bibinfo{volume}{185}},
  \bibinfo{pages}{2250} (\bibinfo{year}{2014}), \eprint{1310.1921}.

\bibitem[{\citenamefont{de~Aquino et~al.}(2011)\citenamefont{de~Aquino,
  Hagiwara, Li, and Maltoni}}]{deAquino:2011ix}
\bibinfo{author}{\bibfnamefont{P.}~\bibnamefont{de~Aquino}},
  \bibinfo{author}{\bibfnamefont{K.}~\bibnamefont{Hagiwara}},
  \bibinfo{author}{\bibfnamefont{Q.}~\bibnamefont{Li}}, \bibnamefont{and}
  \bibinfo{author}{\bibfnamefont{F.}~\bibnamefont{Maltoni}},
  \bibinfo{journal}{JHEP} \textbf{\bibinfo{volume}{06}}, \bibinfo{pages}{132}
  (\bibinfo{year}{2011}), \eprint{1101.5499}.

\bibitem[{\citenamefont{Degrande et~al.}(2012)\citenamefont{Degrande, Duhr,
  Fuks, Grellscheid, Mattelaer, and Reiter}}]{Degrande:2011ua}
\bibinfo{author}{\bibfnamefont{C.}~\bibnamefont{Degrande}},
  \bibinfo{author}{\bibfnamefont{C.}~\bibnamefont{Duhr}},
  \bibinfo{author}{\bibfnamefont{B.}~\bibnamefont{Fuks}},
  \bibinfo{author}{\bibfnamefont{D.}~\bibnamefont{Grellscheid}},
  \bibinfo{author}{\bibfnamefont{O.}~\bibnamefont{Mattelaer}},
  \bibnamefont{and} \bibinfo{author}{\bibfnamefont{T.}~\bibnamefont{Reiter}},
  \bibinfo{journal}{Comput. Phys. Commun.} \textbf{\bibinfo{volume}{183}},
  \bibinfo{pages}{1201} (\bibinfo{year}{2012}), \eprint{1108.2040}.

\bibitem[{\citenamefont{de~Aquino et~al.}(2012)\citenamefont{de~Aquino, Link,
  Maltoni, Mattelaer, and Stelzer}}]{deAquino:2011ub}
\bibinfo{author}{\bibfnamefont{P.}~\bibnamefont{de~Aquino}},
  \bibinfo{author}{\bibfnamefont{W.}~\bibnamefont{Link}},
  \bibinfo{author}{\bibfnamefont{F.}~\bibnamefont{Maltoni}},
  \bibinfo{author}{\bibfnamefont{O.}~\bibnamefont{Mattelaer}},
  \bibnamefont{and} \bibinfo{author}{\bibfnamefont{T.}~\bibnamefont{Stelzer}},
  \bibinfo{journal}{Comput. Phys. Commun.} \textbf{\bibinfo{volume}{183}},
  \bibinfo{pages}{2254} (\bibinfo{year}{2012}), \eprint{1108.2041}.

\bibitem[{\citenamefont{Alwall et~al.}(2014)\citenamefont{Alwall, Frederix,
  Frixione, Hirschi, Maltoni, Mattelaer, Shao, Stelzer, Torrielli, and
  Zaro}}]{Alwall:2014hca}
\bibinfo{author}{\bibfnamefont{J.}~\bibnamefont{Alwall}},
  \bibinfo{author}{\bibfnamefont{R.}~\bibnamefont{Frederix}},
  \bibinfo{author}{\bibfnamefont{S.}~\bibnamefont{Frixione}},
  \bibinfo{author}{\bibfnamefont{V.}~\bibnamefont{Hirschi}},
  \bibinfo{author}{\bibfnamefont{F.}~\bibnamefont{Maltoni}},
  \bibinfo{author}{\bibfnamefont{O.}~\bibnamefont{Mattelaer}},
  \bibinfo{author}{\bibfnamefont{H.~S.} \bibnamefont{Shao}},
  \bibinfo{author}{\bibfnamefont{T.}~\bibnamefont{Stelzer}},
  \bibinfo{author}{\bibfnamefont{P.}~\bibnamefont{Torrielli}},
  \bibnamefont{and} \bibinfo{author}{\bibfnamefont{M.}~\bibnamefont{Zaro}},
  \bibinfo{journal}{JHEP} \textbf{\bibinfo{volume}{07}}, \bibinfo{pages}{079}
  (\bibinfo{year}{2014}), \eprint{1405.0301}.

\bibitem[{\citenamefont{Ball et~al.}(2013)}]{Ball:2012cx}
\bibinfo{author}{\bibfnamefont{R.~D.} \bibnamefont{Ball}} \bibnamefont{et~al.},
  \bibinfo{journal}{Nucl. Phys.} \textbf{\bibinfo{volume}{B867}},
  \bibinfo{pages}{244} (\bibinfo{year}{2013}), \eprint{1207.1303}.

\bibitem[{\citenamefont{Alwall et~al.}(2015)\citenamefont{Alwall, Duhr, Fuks,
  Mattelaer, Özturk, and Shen}}]{Alwall:2014bza}
\bibinfo{author}{\bibfnamefont{J.}~\bibnamefont{Alwall}},
  \bibinfo{author}{\bibfnamefont{C.}~\bibnamefont{Duhr}},
  \bibinfo{author}{\bibfnamefont{B.}~\bibnamefont{Fuks}},
  \bibinfo{author}{\bibfnamefont{O.}~\bibnamefont{Mattelaer}},
  \bibinfo{author}{\bibfnamefont{D.~G.} \bibnamefont{Özturk}},
  \bibnamefont{and} \bibinfo{author}{\bibfnamefont{C.-H.} \bibnamefont{Shen}},
  \bibinfo{journal}{Comput. Phys. Commun.} \textbf{\bibinfo{volume}{197}},
  \bibinfo{pages}{312} (\bibinfo{year}{2015}), \eprint{1402.1178}.

\bibitem[{\citenamefont{Kumar et~al.}(2009)\citenamefont{Kumar, Mathews,
  Ravindran, and Tripathi}}]{Kumar:2009nn}
\bibinfo{author}{\bibfnamefont{M.~C.} \bibnamefont{Kumar}},
  \bibinfo{author}{\bibfnamefont{P.}~\bibnamefont{Mathews}},
  \bibinfo{author}{\bibfnamefont{V.}~\bibnamefont{Ravindran}},
  \bibnamefont{and} \bibinfo{author}{\bibfnamefont{A.}~\bibnamefont{Tripathi}},
  \bibinfo{journal}{Nucl. Phys.} \textbf{\bibinfo{volume}{B818}},
  \bibinfo{pages}{28} (\bibinfo{year}{2009}), \eprint{0902.4894}.

\bibitem[{\citenamefont{Allanach et~al.}(2002)\citenamefont{Allanach, Odagiri,
  Palmer, Parker, Sabetfakhri, and Webber}}]{Allanach:2002gn}
\bibinfo{author}{\bibfnamefont{B.~C.} \bibnamefont{Allanach}},
  \bibinfo{author}{\bibfnamefont{K.}~\bibnamefont{Odagiri}},
  \bibinfo{author}{\bibfnamefont{M.~J.} \bibnamefont{Palmer}},
  \bibinfo{author}{\bibfnamefont{M.~A.} \bibnamefont{Parker}},
  \bibinfo{author}{\bibfnamefont{A.}~\bibnamefont{Sabetfakhri}},
  \bibnamefont{and} \bibinfo{author}{\bibfnamefont{B.~R.}
  \bibnamefont{Webber}}, \bibinfo{journal}{JHEP} \textbf{\bibinfo{volume}{12}},
  \bibinfo{pages}{039} (\bibinfo{year}{2002}), \eprint{hep-ph/0211205}.

\end{thebibliography}

\end{document}